\documentclass{emulateapj}
\usepackage{apjfonts}

\newcommand{\myemail}{liang-n03@mails.tsinghua.edu.cn,xwk@mail.tsinghua.edu.cn,
yuan-liu@mails.tsinghua.edu.cn,zhangsn@mail.tsinghua.edu.cn}

\slugcomment{Accepted for publication in ApJ (Received 2008 February
28; accepted 2008 June 4).}

\shorttitle{Cosmology-Independent GRB Luminosity Relations}
\shortauthors{Liang et al.}

\begin{document}

\title{A Cosmology-Independent Calibration of Gamma-Ray Burst Luminosity Relations  and the Hubble Diagram }

\author{Nan~Liang\altaffilmark{1},  Wei~Ke Xiao\altaffilmark{1},
Yuan~Liu\altaffilmark{1}, Shuang~Nan~Zhang\altaffilmark{1,2,3}}

\altaffiltext{1}{Department of Physics and Tsinghua Center for
Astrophysics, Tsinghua University, Beijing 100084, China; \myemail}

\altaffiltext{2}{Key Laboratory of Particle Astrophysics, Institute
of High Energy Physics, Chinese Academy of Sciences, Beijing 100049,
China}

\altaffiltext{3}{Physics Department, University of Alabama in
Huntsville, Huntsville, AL 35899}

\begin{abstract}
An important concern in the application of gamma-ray bursts (GRBs)
to cosmology is that the calibration of GRB luminosity/energy
relations depends on the cosmological model, due to the lack of a
sufficient low-redshift GRB sample. In this paper, we present a new
method to calibrate GRB relations in a cosmology-independent way.
Since objects at the same redshift should have the same luminosity
distance and since the distance moduli of Type Ia supernovae (SNe
Ia) obtained directly from observations are completely cosmology
independent, we obtain the distance modulus of a GRB at a given
redshift by interpolating from the Hubble diagram of SNe Ia. Then we
calibrate seven GRB relations without assuming a particular
cosmological model and construct a GRB Hubble diagram to constrain
cosmological parameters. From the 42 GRBs at $1.4<z\le6.6$, we
obtain $\Omega_{\rm M}=0.25_{-0.05}^{+0.04}$,
$\Omega_{\Lambda}=0.75_{-0.04}^{+0.05}$ for the flat $\Lambda$CDM
model, and for the dark energy model with a constant equation of
state $w_0=-1.05_{-0.40}^{+0.27}$, which is consistent with the
concordance model in a 1-$\sigma$ confidence region.
\end{abstract}


\section{Introduction}\label{sec1}
In the past decade, observations of Type Ia supernovae (SNe Ia;
Riess et al. 1998; Schmidt et al. 1998; Perlmutter et al. 1999),
cosmic microwave background (CMB) fluctuations (Bennett et al. 2003;
Spergel et al. 2003, 2007), and large-scale structures (LSS; Tegmark
et al. 2004, 2006) have been used to explore cosmology extensively.
These observations are found to be consistent with the so-called
concordance cosmology, in which the universe is spatially flat and
contains pressureless matter and dark energy, with fractional energy
densities of $\Omega_{\rm M}=0.27\pm0.04$ and
$\Omega_{\Lambda}=0.73\pm0.04$ (Davis et al. 2007). Observations of
SN Ia provide a powerful probe in the modern cosmology. Phillips
(1993) found that there is an intrinsic relation between the peak
luminosity and the shape of the light curve of SNe Ia. This relation
and other similar luminosity relations make SNe Ia standard candles
for measuring the geometry and dynamics of the universe. However,
the maximum redshift of the SNe Ia which we can currently use is
only about 1.7, whereas fluctuations of the CMB provide the
cosmological information from last scattering surface at $z=1089$.
Therefore, the earlier universe at higher redshift may not be
well-studied without data from standardized candles in the
``cosmological desert'' from the SNe~Ia redshift limit to $z\sim
1000$.

Gamma-Ray Bursts (GRBs) are the most intense explosions observed so
far. Their high energy photons in the gamma-ray band are almost
immune to dust extinction, whereas in the case of SN Ia
observations, there is extinction from the interstellar medium when
optical photons propagate towards us. Moreover, GRBs are likely to
occur in the high-redshift range up to at least $z=6.6$ (Krimm et
al. 2006); higher redshift GRBs up to $z=10$ should have already
been detected, although none have been identified (Lamb \& Reichart
2000; Bromm \& Loeb 2002, 2006; Lin, Zhang, \& Li 2004). Thus, by
using GRBs, we may explore the early universe in the high redshift
range which is difficult to access by other cosmological probes.
These advantages make GRBs attractive for cosmology research.

GRB luminosity/energy relations are connections between measurable
properties of the prompt gamma-ray emission and the luminosity or
energy. In recent years, several empirical GRB luminosity relations
have been proposed as distance indicators (see e.g.  Ghirlanda et
al. 2006a; Schaefer (2007) for reviews), such as the
luminosity-spectral lag ($L$-$\tau_{\rm lag}$) relation (Norris,
Marani, \& Bonnell 2000), the luminosity-variability ($L$-$V$)
relation (Fenimore \& Ramirez-Ruiz 2000; Reichart et al. 2001), the
isotropic energy-peak spectral energy (($E_{\rm iso}$-$E_{\rm p}$)
relation (i.e., the so-called Amati relation, Amati et al. 2002),
the collimation-corrected energy-peak spectral energy
($E_{\gamma}$-$E_{\rm p}$) relation (i.e., the so-called Ghirlanda
et al. 2004a), the $L$-$E_{\rm p}$ relation (Schaefer 2003a;
Yonetoku et al. 2004), and two multi-variable relations. The first
of these multiple relations is between $E_{\rm iso}$, $E_{\rm p}$
and the break time of the optical afterglow light curves $(t_{\rm
b})$ (i.e., the so-called Liang-Zhang relation, Liang \& Zhang
2005); the other is between luminosity, $E_{\rm p}$ and the
rest-frame ``high-signal'' timescale ($T_{0.45}$) (Firmani et al.
2006).

Several authors have made use of these GRB luminosity indicators as
standard candles at very high redshift for cosmology research (e.g.
Schaefer 2003b; Dai et al. 2004; Ghirlanda et al. 2004b; Firmani et
al. 2005, 2006, 2007; Liang \& Zhang 2005; Xu et al. 2005; Wang \&
Dai 2006; see also e.g. Ghirlanda et al. 2006a and Schaefer (2007)
for reviews). Schaefer (2003b) derived the luminosity distances of
nine GRBs with known redshifts by using two GRB luminosity relations
to construct the first GRB Hubble diagram. Dai et al. (2004)
considered the Ghirlanda relation with 12 bursts and proposed
another approach to constrain cosmological parameters. Liang \&
Zhang (2005) constrained cosmological parameters and the transition
redshift using the $E_{\rm iso}$-$E_{\rm p}$-$t_{\rm b}$ relation.
More recently, Schaefer (2007) used five GRB relations calibrated
with 69 GRBs by assuming two adopted cosmological models to obtain
the derived distance moduli for plotting the Hubble diagram,  and
joint constraints on the cosmological parameters and dark energy
models have been derived in many works by combining the 69 GRB data
with SNe Ia and the other cosmological probes, such as the CMB
anisotropy, the baryon acoustic oscillation (BAO) peak, the X-ray
gas mass fraction in clusters, the linear growth rate of
perturbations, and the angular diameter distances to radio galaxies
(Wright 2007; Wang et al. 2007; Li et al. 2008;  Qi et al. 2008;
Daly et al. 2008).

However, an important point related to the use of GRBs for cosmology
is the dependence on the cosmological model in the calibration of
GRB relations. In the case of SN Ia cosmology, the calibration is
carried out with a sample of SNe Ia at very low redshift where the
luminosities of SNe Ia are essentially independent of any
cosmological model (i.e., at $z\le0.1$, the luminosity distance has
a negligible dependence on the choice of the model). However, in the
case of GRBs, the observed long-GRB rate falls off rapidly at low
redshifts, and some nearby GRBs may be intrinsically different
(e.g., GRB 980425, GRB 031203; Norris 2002; Soderberg et al. 2004;
Guetta et al. 2004; Liang \& Zhang 2006a). Therefore, it is very
difficult to calibrate the relations with a low-redshift sample. The
relations of GRBs presented above have been calibrated by assuming a
particular cosmological model (e.g. the $\Lambda$CDM model). In
order to investigate cosmology, the relations of standard candles
should be calibrated in a cosmological model-independent way.
Otherwise, the circularity problem cannot easily be avoided. Many of
the works mentioned above treat the circularity problem with a
statistical approach.  A simultaneous fit of the parameters in the
calibration curves and the cosmology is carried out to find the
optimal GRB relation and the optimal cosmological model in the sense
of a minimum scattering in both the luminosity relations and the
Hubble diagram. Firmani et al. (2005) has also proposed a Bayesian
method to get around the circularity problem. Li et al. (2008b)
presented another new method to deal with the problem, using Markov
chain Monte Carlo (MCMC) global fitting analysis. Instead of a
hybrid sample over the whole redshift range of GRBs, Takahashi et
al. (2003) first calibrated two GRB relations at low redshift
($z\le1$), where distance-redshift relations have been already
determined from SN Ia; they then used GRBs at high redshift ($z>1$)
as a distance indicator. Bertolami \& Silva (2006) considered the
use of GRBs at $1.5<z<5$ calibrated with the bursts at $z\le1.5$ as
distance markers to study the unification of dark energy and dark
matter in the context of the generalized Chaplygin gas model.
Schaefer (2007) made a detailed comparison between the Hubble
diagram calibrated only with 37 bursts at $z<2$ and the one
calibrated with the data from all 69 GRBs to show that the results
from  fits to the GRB Hubble diagram calibrated with the hybrid
sample are robust. These calibration methods above carried out with
the sample at low redshifts  have been derived from the $\Lambda$CDM
model. However, we note that the circularity problem can not be
circumvented completely by means of statistical approaches, because
a particular cosmology model is required in doing the joint fitting.
This means that the parameters of the calibrated relations are still
coupled to the cosmological parameters derived from a given
cosmological model. In principle, the circularity problem can be
avoided in two ways (Ghirlanda et al. 2006a): (1) through a solid
physical interpretation of these relations that would fix their
slope independently from cosmological model, or (2) through the
calibration of these relations by several low-redshift GRBs.
Recently the possibility of calibrating the standard candles using
GRBs in a low-dispersion in redshift near a fiducial redshift has
been proposed (Lamb et al. 2005; Ghirlanda et al. 2006b). Liang \&
Zhang (2006b) elaborated this method further based on  Bayesian
theory. However, the GRB sample available now is far from what is
needed to calibrate the relations in this way.

As analyzed above, due to the lack of low-redshift GRBs, the methods
used in GRB luminosity relations for cosmology are different from
the standard Hubble diagram method used in SN Ia cosmology. In this
work, we present a new method to calibrate the GRB relations in a
cosmological model-independent way. In the case of SN Ia cosmology,
the distance of nearby SNe Ia used to calibrate the luminosity
relations can be obtained by measuring Cepheid variables in the same
galaxy, and with other distance indicators. Thus, Cepheid variables
have been regarded as the first-order standard candles for
calibrating SNe Ia, the second-order standard candles. It is obvious
that objects at the same redshift should have the same luminosity
distance in any cosmology. For calibrating GRBs, there are so many
SNe Ia (e.g., 192 SNe Ia used in Davis et al. 2007) that we can
obtain the distance moduli (also the luminosity distance) at any
redshift in the redshift range of SNe Ia by interpolating from SN Ia
data in the Hubble diagram. Furthermore, the distance moduli of SNe
Ia obtained directly from observations are completely cosmological
model independent. Therefore, in the same sense as Cepheid variables
and SNe Ia, if we regard SNe Ia as the first order standard candles,
we can obtain the distance moduli of GRBs in the redshift range of
SNe Ia and calibrate GRB relations in a completely cosmological
model-independent way. Then if we further assume that these
calibrated GRB relations  are still valid in the whole redshift
range of GRBs, just as for SNe Ia, we can use the standard Hubble
diagram method to constrain the cosmological parameters from the GRB
data at high redshift obtained by utilizing the relations.

The structure of this paper is arranged as follow. In section 2, we
calibrate seven GRB luminosity/energy relations with  the sample at
$z\le1.4$ obtained by interpolating from SNe Ia data in the Hubble
diagram. In section 3, we construct the Hubble diagram of GRBs
obtained by using the interpolation methods and constrain
cosmological parameters. Conclusions and a discussion are given in
section 4.

\section{The Calibration of the Luminosity Relations of Gamma-Ray Bursts}
We adopt the data for 192 SNe Ia (Riess et al. 2007; Wood-Vasey et
al. 2007; Astier et al. 2007, Davis et al. 2007), shown in Figure 1.
It is clear that for GRBs in the redshift range of SNe Ia, there are
enough data SN Ia points  and the redshift intervals of the
neighboring SN Ia data points are also small enough, to be able to
use interpolation methods to obtain the distance moduli of GRBs.
Since there is only one SN Ia point at $z>1.4$ (the redshift of
SN1997ff is $z=1.755$), we initially exclude it from our SN Ia
sample used to interpolate the distance moduli of GRBs in the
redshift range of the SN Ia sample (191 SNe Ia data at $z\le1.4$).
As the distance moduli of SNe Ia data in the Hubble diagram are
obtained directly from observations, with the sample at $z\le1.4$ by
interpolating from the Hubble diagram of SNe Ia, we can therefore
calibrate GRB luminosity relations in a completely
cosmology-independent way.

In this section, we calibrate seven GRB luminosity/energy relations
with the sample at $z\le1.4$, i.e., the $\tau_{\rm lag}$-$L$
relation, the $V$-$L$ relation, the $L$-$E_{\rm p}$ relation, the
$E_{\gamma}$-$E_{\rm p}$ relation, the $\tau_{\rm RT}$-$L$ relation,
where $\tau_{\rm RT}$ is the minimum rise time in the GRB light
curve (Schaefer, 2007), the $E_{\rm iso}$-$E_{\rm p}$ relation, and
the $E_{\rm iso}$-$E_{\rm p}$-$t_{\rm b}$ relation.

The isotropic luminosity of a burst is calculated  by
\begin{equation}
L = 4\pi d_{\rm L}^2 P_{\rm bolo},
\end{equation}
where $d_{\rm L}$ is the luminosity distance of the burst and
$P_{\rm bolo}$ is the bolometric flux of gamma-rays in the burst.
The isotropic energy released from a burst is given by
\begin{equation}
E_{\rm iso} = 4\pi d_{\rm L}^2 S_{\rm bolo} (1+z)^{-1},
\end{equation}
where $S_{\rm bolo}$ is the bolometric fluence of gamma-rays in the
burst at redshift $z$. The total collimation-corrected energy is
then calculated by
\begin{equation}
E_{\gamma} = F_{\rm beam}E_{\rm iso},
\end{equation}
where the beaming factor, $F_{\rm beam}$ is $(1-\cos\theta _{\rm
jet})$ with the jet opening angle ($\theta_{\rm jet}$), which is
related to the break time ($t_{\rm b}$).

A GRB luminosity relation can be generally written in the form
\begin{equation}
\log y=a+b\log x,
\end{equation}
where $a$ and $b$ are the intercept and slope of the relation
respectively; $y$ is the luminosity ($L$ in units of $\rm erg\
s^{-1}$) or energy ($E_{\ \rm iso}$ or $E_{\gamma}$ in units of $\rm
erg$); $x$ is the GRB parameters measured in the rest frame, e.g.,
$\tau_{\rm lag}(1+z)^{-1}/(0.1\rm \ s)$, $V(1+z)/0.02$, $E_{\rm
p}(1+z)/(300\ \rm keV)$, $\tau_{\rm RT}(1+z)^{-1}/\rm (0.1\ s)$,
$E_{\rm p}(1+z)/(300\ \rm keV)$, for the 6 two-variable relations
above. We adopt the data for these quantities from Schaefer (2007).
For the one multi-variable relation (i.e., $E_{\rm iso}$-$E_{\rm
p}$-$t_{\rm b}$), the calibration equation is
\begin{equation}
\log y=a +b_1\log x_1 + b_2\log x_2,
\end{equation}
where $x_1$ and $x_2$ are $E_{\rm p}(1+z)/(300\ {\rm keV)}$, $t_{\rm
b}/(1+z)/(1 {\ \rm day)}$ respectively, and  $b_1$ and $b_2$ are the
slopes of $x_1$ and $x_2$ respectively. The error of the
interpolated distance modulus of a GRB must account for the original
uncertainties of the SNe Ia as well as for the uncertainties from
the interpolation. When the linear interpolation is used, the error
can be calculated by
\begin{equation}
\sigma_{\mu}=([(z_{i+1}-z)/(z_{i+1}-z_i)]^2\epsilon_{\mu,i}^2+[(z-z_{i})/(z_{i+1}-z_i)]^2\epsilon_{\mu,i+1}^2)^{1/2},
\end{equation}
where  $\sigma_{\mu}$ is the error of the interpolated distance
modulus, $\mu$ is the interpolated distance modulus of a source at
redshift $z$, $\epsilon_{\mu,i}$ and $\epsilon_{\mu,i+1}$ are errors
of the SNe, $\mu_{i}$ and  $\mu_{i+1}$ are the distance moduli of
the SNe at nearby redshifts $z_{i}$ and $z_{i+1}$ respectively. When
the cubic interpolation is used, the error can be calculated by
\begin{equation}
\sigma_{\mu}=(A_0^2\epsilon_{\mu,i}^2+A_1^2\epsilon_{\mu,i+1}^2+A_2^2\epsilon_{\mu,i+2}^2+A_3^2\epsilon_{\mu,i+3}^2)^{1/2},
\end{equation}
  where $\epsilon_{\mu,i}$, $\epsilon_{\mu,i+1}$,
$\epsilon_{\mu,i+2}$, and $\epsilon_{\mu,i+3}$  are errors of the
SNe; and $\mu_{i}$, $\mu_{i+1}$, $\mu_{i+2}$, and $\mu_{i+3}$  are
the distance moduli of the SNe at nearby redshifts $z_{i}$,
$z_{i+1}$, $z_{i+2}$, and $z_{i+3}$:

$A_0=[(z_{i+1}-z)(z_{i+2}-z)(z_{i+3}-z)]/[(z_{i+1}-z_{i})(z_{i+2}-z_{i})(z_{i+3}-z_{i})]$;
$A_1=[(z_{i}-z)(z_{i+2}-z)(z_{i+3}-z)]/[(z_{i}-z_{i+1})(z_{i+2}-z_{i+1})(z_{i+3}-z_{i+1})]$;
$A_2=[(z_{i}-z)(z_{i+1}-z)(z_{i+3}-z)]/[(z_{i}-z_{i+2})(z_{i+1}-z_{i+2})(z_{i+3}-z_{i+2})]$;
$A_3=[(z_{i}-z)(z_{i+1}-z)(z_{i+2}-z)]/[(z_{i}-z_{i+3})(z_{i+1}-z_{i+3})(z_{i+2}-z_{i+3})]$.

\begin{figure}
\includegraphics[angle=0,width=0.47\textwidth]{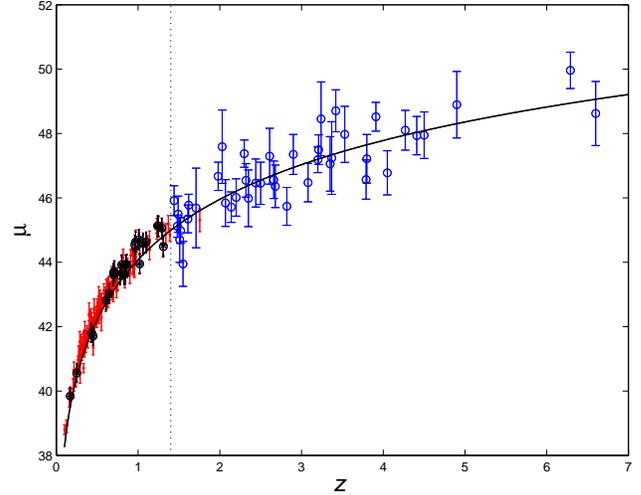}
\caption{Hubble Diagram of 192 SNe Ia (\emph{red dots}) and the 69
GBRs (\emph{circles}) obtained  using the interpolation methods. The
27 GBRs at $z\le1.4$ are obtained by interpolating from SN Ia data
(\emph{black circles}: the cubic interpolation method; \emph{black
stars}: the linear interpolation method); and the 42 GBRs at $z>1.4$
(\emph{blue circles}) are obtained with the five relations
calibrated with the sample at $z\le1.4$ using the cubic
interpolation method. The curve is the theoretical distance modulus
in the concordance model ($w_0= -1$, $\Omega_{\rm M}= 0.27$,
$\Omega_{\Lambda}= 0.73$), and the vertical dotted line represents
$z=1.4$. } \label{fig1}

\end{figure}

We determine the values of the intercept ($a$) and the slope ($b$)
with their 1-$\sigma$ uncertainties calibrated with the GRB sample
at $z\le1.4$ by using two interpolation methods (the linear
interpolation methods and the cubic interpolation method). For the 6
two-variable relations we use the same method (the bisector of the
two ordinary least-squares) as used in Schaefer (2007), and for the
one multi-variable relation, the multiple variable regression
analysis is used. The bisector of the two ordinary least-squares
(Isobe et al., 1990) does not take the errors into account; but the
use of weighted least-squares, taking into account the measurable
uncertainties, results in almost identical best fits. However,
taking into account the  measurable uncertainties in the regression,
 when they are smaller than the intrinsic error, indeed
will not change the fitting parameters significantly (for further
discussion, see Schaefer 2007). The calibration results are
summarized in Table 1, and we plot the GRB data at $z\le1.4$ with
the distance moduli obtained by using the two interpolation methods
from SN Ia data in Figure 1 for comparison. In the previous
treatment, the distance moduli of GRBs are obtained by assuming a
particular cosmological model with a hybrid sample in the whole
redshift range of GRBs to calibrate the relations. Therefore, for
comparison in Table 1  we also list the  results calibrated with the
same sample ($z\le1.4$) assuming the $\Lambda$CDM model ($w=-1$) or
the Riess cosmology\footnote{Here we follow Schaefer (2007) in
calling this particular parametrization of the equation of state the
"Riess cosmology".} ($w=-1.31+1.48z$, Riess et al. 2004), which were
used to calibrate the five GRB relations in Schaefer (2007). The
calibration results obtained by using the interpolation methods are
carried out with the 27 GRBs at $z\le1.4$, namely, 13, 19, 25, 12,
24, 12, and 12 GRBs for the $\tau_{\rm lag}$-$L$, $V$-$L$,
$L$-$E_{\rm p}$, $E_{\gamma}$-$E_{\rm p}$, $\tau_{\rm RT}$-$L$,
$E_{\rm iso}$-$E_{\rm p}$, and $E_{\rm iso}$-$E_{\rm p}$-$t_{\rm b}$
relations respectively;\footnote{For the seven relations, only data
from Table 4 in Schaefer (2007) are used for the calibration.
However, we note that there are more GRB data available to calibrate
the $E_{\rm iso}$-$E_{\rm p}$ relation.} therefore, the
uncertainties of the results are somewhat larger than those with the
hybrid GRB sample.\footnote{Calibration results with their
1-$\sigma$ uncertainty for the five GRB relations with 69 GRBs
($0.1<z\le6.6$) obtained by assuming the $\Lambda$CDM model in
Schaefer (2007) are also given in Table 1, for comparison.}

\begin{table*}
\begin{center}
  \caption{Calibration results} {\scriptsize

\begin{tabular}{lccccccccccc}

  \hline\hline
 & \multicolumn{2}{c}{Linear interpolation method}
 & \multicolumn{2}{c}{Cubic interpolation method}
 & \multicolumn{2}{c}{The $\Lambda$CDM
model}
 & \multicolumn{2}{c}{The Riess cosmology}
 & \multicolumn{2}{c}{Schaefer
(2007)} & \\
\cline{2-3} \cline{4-5} \cline{6-7} \cline{8-9}
\cline{10-11}Relation &
a&b&   a&b&   a&b&   a&b &  a&b \\
 \hline
$\tau_{\rm lag}$-$L$                                    &52.22$\pm0.09$& -1.07$\pm0.14$&52.22$\pm0.09$& -1.07$\pm0.13$&52.15$\pm0.10$& -1.11$\pm0.14$&52.13$\pm0.09$& -1.10$\pm0.13$ &52.26$\pm0.06$& -1.01$\pm0.05$\\
$V$-$L$                                                 &52.58$\pm0.13$&  2.04$\pm0.26$&52.59$\pm0.13$&  2.05$\pm0.27$&52.48$\pm0.13$&  2.04$\pm0.30$&52.47$\pm0.13$&  2.02$\pm0.30$ &52.49$\pm0.22$&  1.77$\pm0.12$\\
$L$-$E_{\rm p}$                                         &52.25$\pm0.09$&  1.68$\pm0.11$&52.26$\pm0.09$&  1.69$\pm0.11$&52.15$\pm0.09$&  1.66$\pm0.12$&52.15$\pm0.09$&  1.64$\pm0.12$ &52.21$\pm0.13$&  1.68$\pm0.05$\\
$E_{\gamma}$-$E_{\rm p}$                                &50.70$\pm0.06$&  1.77$\pm0.17$&50.71$\pm0.07$&  1.79$\pm0.18$&50.59$\pm0.06$&  1.75$\pm0.19$&50.59$\pm0.06$&  1.73$\pm0.19$ &50.57$\pm0.09$&  1.63$\pm0.03$\\
$\tau_{\rm RT}$-$L$                                     &52.63$\pm0.11$& -1.31$\pm0.15$&52.64$\pm0.11$& -1.31$\pm0.15$&52.52$\pm0.10$& -1.30$\pm0.15$&52.51$\pm0.10$& -1.29$\pm0.15$ &52.54$\pm0.06$& -1.21$\pm0.06$ \\
$E_{\rm iso}$-$E_{\rm p}$                               &52.99$\pm0.16$&  1.88$\pm0.23$&53.00$\pm0.17$&  1.90$\pm0.23$&52.88$\pm0.16$&  1.84$\pm0.21$&52.88$\pm0.15$&  1.81$\pm0.20$ \\
$E_{\rm iso}$-$E_{\rm p}$-$t_{\rm b}$ \tablenotemark{a}&52.83$\pm0.10$&  2.26$\pm0.30$&52.83$\pm0.10$&  2.28$\pm0.30$&52.73$\pm0.10$&  2.19$\pm0.30$&52.73$\pm0.10$&  2.16$\pm0.30$ \\
                                                                                 &              & -1.07$\pm0.21$&              & -1.07$\pm0.21$&              & -1.01$\pm0.20$&              & -0.99$\pm0.21$ \\
 \hline
\end{tabular}
}
\parbox{0.9\textwidth}{
\tablecomments{Calibration results (for $a$=intercept,  $b$=slope)
with their 1-$\sigma$ uncertainties, for the seven GRB
luminosity/energy relations with the sample at $z\le1.4$, using two
interpolation methods (the linear interpolation method and the cubic
interpolation methods) directly from SNe Ia data and by assuming a
particular cosmological models (the $\Lambda$CDM model or the Riess
cosmology). Only the data available from Table 4 in Schaefer 2007
are used to calibrate the seven GRB relations. For comparison,
calibration results with the 69 GRBs obtained  by assuming the
$\Lambda$CDM model for the five GRB relations in Schaefer 2007 are
given in the last column.}
 \tablenotetext{a}{In the slope values for
the $E_{\rm iso}$-$E_{\rm p}$-$t_{\rm b}$ relation, the first line
is $b_1$, and the second line is $b_2$.}}
\end{center}
\end{table*}

\clearpage

The linear correlation coefficients of the calibration with the
sample at $z\le1.4$ using the cubic interpolation methods are -0.88,
0.65, 0.89, 0.94, -0.75, and 0.68 for the above 6 two-variable
relations respectively, and 0.94 for $E_{\rm iso}$ vs. the combined
variable ($b_1\log[E_{\rm p}(1+z)] + b_2\log[t_{\rm b}/(1+z)]$) for
the $E_{\rm iso}$-$E_{\rm p}$-$t_{\rm b}$ relation, which shows that
these correlations are significant.

From Table 1 and Figure 1, we find that the calibration results
obtained using the linear interpolation methods  are almost
identical to the results calibrated by using the cubic interpolation
method. We also find the results obtained by assuming the two
cosmological models with the same sample differ only slightly from,
but are still fully consistent with, those calibrated using our
interpolation methods. The reason for this is easy to understand,
since both cosmological models are fully compatible with SN Ia data.
Nevertheless, it should be noticed that the calibration results
obtained using the interpolation methods directly from SN Ia data
are completely cosmology independent.

\section{The Hubble Diagram of Gamma-Ray Bursts}
If we further assum GRB luminosity/energy relations do not evolve
with redshift, we are able to obtain the luminosity ($L$) or energy
($E_{\rm iso}$ or $E_\gamma$) of each burst at high redshift
($z>1.4$) by utilizing the calibrated relations. Therefore, the
luminosity distance ($d_L$) can be derived from  equation
(1$\sim$3). The uncertainty of the value of the luminosity or energy
deduced from a GRB relation is
\begin{equation}
\sigma_{\log y}^2 = \sigma_a^2 + (\sigma_b \log x)^2 + (0.4343 b
\sigma_x / x)^2 + \sigma_{\rm sys}^2,
\end{equation}
where $\sigma_{a}$, $\sigma_b$ and  $\sigma_{x}$ are 1-$\sigma$
uncertainty of the intercept, the slope and the GRB measurable
parameters, and $\sigma_{\rm sys}$ is the systematic error in the
fitting that accounts for the extra scatter of the luminosity
relations. The value of $\sigma_{\rm sys}$ can be estimated by
finding the value such that a $\chi ^2$  fit to the calibration
curve produces a value of reduced $\chi ^2$ of unity (Schaefer
2007). A distance modulus can be calculated as $\mu=5\log d_L + 25$
(where $d_L$ is in Mpc). The propagated uncertainties will depend on
whether $P_{\rm bolo}$ or $S_{\rm bolo}$ is used:
\begin{equation}
\sigma_{\mu} = [(2.5 \sigma_{\log L})^2+(1.086 \sigma_{P_{\rm
bolo}}/P_{\rm bolo})^2]^{1/2},
\end{equation}
or
\begin{equation}
\sigma_{\mu} = [(2.5 \sigma_{\log E_{\rm iso}})^2+(1.086
\sigma_{S_{\rm bolo}}/S_{\rm bolo})^2]^{1/2},
\end{equation}
and
\begin{equation}
\sigma_{\mu} = [(2.5 \sigma_{\log E_{\gamma}})^2+(1.086
\sigma_{S_{\rm bolo}}/S_{\rm bolo})^2+(1.086 \sigma_{F_{\rm
beam}}/F_{\rm beam})^2]^{1/2}.
\end{equation}

We use the same method as used in Schaefer (2007) to obtain the best
estimate $\mu$ for each GRB which is the weighted average of all
available distance moduli. The derived distance modulus for each GRB
is
\begin{equation}
\mu = (\sum_i \mu_{\rm i} / \sigma_{\mu_{\rm i}}^2)/(\sum_i
\sigma_{\mu_{\rm i}}^{-2}),
\end{equation}
with its uncertainty $ \sigma_{\mu} = (\sum_i \sigma_{\mu_{\rm
i}}^{-2})^{-1/2}$, where the summations run from 1 to 5  over the
five relations used in Schaefer (2007) with available data.

We have plotted the Hubble diagram of the 69 GRBs obtained  using
the interpolation methods in Figure 1. The 27 GRBs at $z\le1.4$ are
obtained by using two interpolation methods directly from SNe data.
The 42 GRB data at $z>1.4$ are obtained by utilizing the five
relations calibrated  with the sample at $z\le1.4$ using the cubic
interpolation method.

Then the cosmological parameters can be fitted by the minimum
$\chi^2$ method. The definition of $\chi^2$ is
\begin{equation}
\chi ^2  = \sum\limits_{i = 1}^N {\frac{{(\mu _{{\rm th},\;i}  - \mu
_{{\rm obs},\;i} )^2 }}{{{\sigma_{\mu_{\rm i}}^2 }}}},
\end{equation}
in which $\mu_{{\rm th},\;i}$ is the theoretical value of distance
modulus, $\mu _{{\rm obs},\;i}$ is the observed distance modulus
with
 its error $\sigma_{\mu_{\rm i}}$. The theoretical value of
the distance modulus ($\mu_{\rm th}$)  depends on the theoretical
value of luminosity distance. For the $\Lambda$CDM model, the
luminosity distance is calculated by
\begin{equation}
 d_L  = \frac{{c(1 + z)}}{{H_0 \sqrt { \left| {\Omega _{\rm k} } \right|} }} \times  \\
 {\mathop{\rm sinn}\nolimits} [\sqrt {\left| {\Omega _{\rm k} } \right|} \int_0^z {\frac{{dz}}{{\sqrt {(1 + z)^2 (1 + \Omega _{\rm m} z) - z(2 + z)\Omega _\Lambda  } }}]} ,  \\
 \end{equation}
where $\Omega _{\rm k}=1-\Omega _{\rm m}-\Omega _\Lambda$,  and
$\rm{ sinn}$$(x)$ is $\rm sinh$ for $\Omega _{\rm k}>0$, $\rm sin$
for $\Omega _{\rm k}<0$, and $x$ for $\Omega _{\rm k}=0$. For the
dark energy model with a constant equation of state ($w_0$), the
luminosity distance in a flat universe is
\begin{equation}
 d_L  = \frac{{c(1 + z)}}{{H_0}} \times\\
\int_0^z {\frac{{dz}}{{\sqrt {(1 + z)^3 \Omega _{\rm m} + (1-\Omega _{\rm m})(1 + z)^{3(1+w_0)}}}}} ,  \\
 \end{equation}
here we adopt $H_0=70\  \rm km\ s^{-1}Mpc^{-1}$.

Figure 2$a$ shows the joint confidence regions for ($\Omega_{\rm
M},\Omega_{\Lambda}$) in the $\Lambda$CDM model from the 42 GRB data
($z>1.4$) obtained by utilizing the five relations calibrated with
the sample at $z\le1.4$ using the cubic interpolation method. The
1-$\sigma$ confidence region is $(\Omega_{\rm
M},\Omega_{\Lambda})=(0.25_{-0.08}^{+0.07},0.68_{-0.64}^{+0.37})$
with $\chi_{\rm min}^2 =44.40$ for 40 degrees of freedom. For a flat
universe, we obtain $\Omega_{\rm M}=0.25_{-0.05}^{+0.04}$ and
$\Omega_{\Lambda}=0.75_{-0.04}^{+0.05}$, which is consistent with
the concordance model ($\Omega_{\rm M}=0.27$,
$\Omega_{\Lambda}=0.73$) in the 1-$\sigma$ confidence region. From
equation (14), it is obvious that for the $\Lambda$CDM model, the
theoretical value of the luminosity distance mainly depends on
$\Omega_{\rm M}$ at higher redshift, and strongly depends on
$\Omega_{\Lambda}$ at lower redshift. Therefore, we can find that
the shape of the likelihood contour of GRBs at higher redshift is
almost vertical to the horizontal axis ($\Omega_{\rm M}$) compared
to that of SNe Ia at lower redshift. To understand the shape of the
confidence regions in the ($\Omega_{\rm M},\Omega_{\Lambda}$) plane,
Firmani et al. (2007)  explored the behavior of the luminosity
distance $d_{\rm L}$ at different redshifts $z$ in a given
cosmological parameter space. The stripe where $d_{\rm L}$ varies by
$1\%$ at $z=3$, shown in the ($\Omega_{\rm M},\Omega_{\Lambda}$)
plane, was almost vertical to the $\Omega_{\rm M}$ axis, which
corresponds roughly to the typical redshifts ($z\sim3$) of the GRB
sample at $z>1.4$. If we add the one SN Ia point (SN1997ff at
$z=1.755$) at $z>1.4$ into our SN Ia sample used to interpolate the
distance moduli of GRBs, we can calibrate the GRB relations with 36
GRBs at $z\le1.76$ by using the cubic interpolation method. We find
that the fitting results from the 33 GRB data ($z>1.76$) obtained by
utilizing the relations calibrated with the sample at $z\le1.76$
using the cubic interpolation method is $\Omega_{\rm
M}=0.28_{-0.07}^{+0.06}$ for a flat universe (Figure 2$b$), which is
consistent with the result from the 42 GRB data ($z>1.4$) obtained
using the cubic interpolation method. However, if the relations are
calibrated with the GRB sample at $z\le1.4$ by assuming a particular
cosmological model, we find that the fitting results  the 42 GRB
data ($z>1.4$) obtained  assuming the $\Lambda$CDM model or the
Riess cosmology are $\Omega_{\rm M}=0.38_{-0.07}^{+0.07}$ (Figure
2$c$) or $\Omega_{\rm M}=0.39_{-0.07}^{+0.07}$ (Figure 2$d$) for a
flat universe, which are systematically higher than $\Omega_{\rm
M}=0.27_{-0.04}^{+0.04}$ beyond a 1-$\sigma$ deviation.

\begin{figure}
  \includegraphics[angle=0,width=0.47\textwidth]{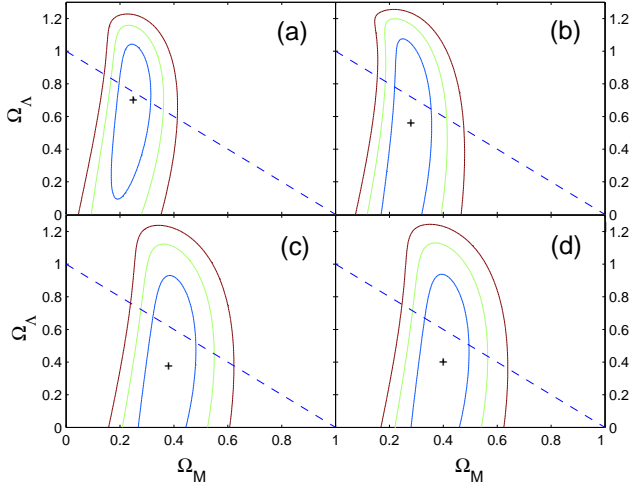}
\caption{(\emph{a}) Joint confidence regions for ($\Omega_{\rm M},
\Omega_{\Lambda}$) in the $\Lambda$CDM model from the data for 42
GRBs ($z>1.4$) obtained by utilizing the five relations calibrated
with the sample at $z\le1.4$ using the cubic interpolation method.
The plus sign indicates the best fit values. The contours correspond
to 1, 2, and 3-$\sigma$ confidence regions, and the dashed line
represents the flat universe. The 1-$\sigma$ confidence region are
$(\Omega_{\rm
M},\Omega_{\Lambda})=(0.25_{-0.08}^{+0.07},0.68_{-0.64}^{+0.37})$.
For a flat universe prior, $\Omega_{\rm M}=0.25_{-0.05}^{+0.04}$ and
$\Omega_{\Lambda}=0.75_{-0.04}^{+0.05}$. (\emph{b}) Joint confidence
regions for ($\Omega_{\rm M}, \Omega_{\Lambda}$) from the data for
33 GRBs ($z>1.76$) obtained by utilizing the relations calibrated
with the sample at $z\le1.76$ by using the cubic interpolation
method. For a flat universe prior, $\Omega_{\rm
M}=0.28_{-0.07}^{+0.06}$. (\emph{c}) The joint confidence regions
for ($\Omega_{\rm M}, \Omega_{\Lambda}$) from the data for 42 GRBs
($z>1.4$) obtained by utilizing the relations calibrated  with the
sample at $z\le1.4$, assuming a $\Lambda$CDM model. For a flat
universe prior, $\Omega_{\rm M}=0.38_{-0.07}^{+0.07}$. (\emph{d})
The joint confidence regions for ($\Omega_{\rm M},
\Omega_{\Lambda}$) from the 42 GRBs data ($z>1.4$) obtained by
utilizing the relations calibrated  with the sample at $z\le1.4$,
assuming a Riess cosmology. For a flat universe prior, $\Omega_{\rm
M}=0.39_{-0.07}^{+0.07}$.} \label{fig2}
\end{figure}

Figure 3 shows likelihood contours in the ($\Omega_{\rm M}, w_0$)
plane in the dark energy model with a constant $w_0$ for a flat
universe from the 42 GRB data ($z>1.4$) obtained by utilizing the
five relations calibrated  with the sample at $z\le1.4$  using the
cubic interpolation method. The best-fit values are $\Omega_{\rm
M}=0.25, w_0=-0.95$. For a prior of $\Omega_{\rm M}=0.27$, we obtain
$w_0=-1.05_{-0.40}^{+0.27}$, which is consistent with the
cosmological constant in a 1-$\sigma$ confidence region.

\begin{figure}
  \includegraphics[angle=0,width=0.47\textwidth]{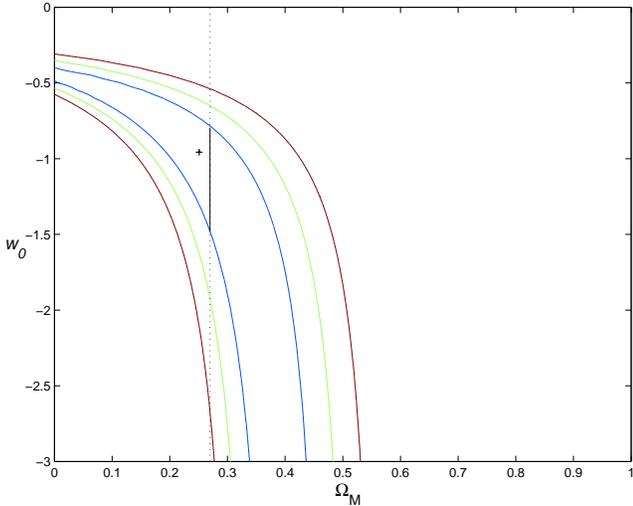}
\caption{Contours of likelihood in the ($\Omega_{\rm M}, w_0$) plane
in the dark energy model with a constant $w_0$ for a flat universe
from the 42 GRBs data ($z>1.4$) obtained by utilizing the five
relations calibrated  with the sample at $z\le1.4$ using the cubic
interpolation method. The plus sign indicates the best-fit values
($\Omega_{\rm M}=0.25, w_0=-0.95$). The contours correspond to 1, 2,
and 3-$\sigma$ confidence regions. For a prior of $\Omega_{\rm
M}=0.27$ (\emph{vertical line}), $w_0=-1.05_{-0.27}^{+0.40}$
(\emph{solid part of vertical  line}).} \label{fig3}
\end{figure}

\section{Summary and Discussion}

With the basic assumption that objects at the same redshift should
have the same luminosity distance, we can obtain the distance
modulus of a GRB at given redshift by interpolating from the Hubble
diagram of SNe Ia at $z\le1.4$. Since the distance modulus of SN Ia
is completely cosmological model independent, the GRB luminosity
relations can be calibrated in a completely cosmology independent
way. Instead of a hybrid sample in the whole redshift range of GRBs
used in most previous treatment, we choose the GRB sample only in
the redshift range of SNe Ia to calibrate the relations. We find
there are not significant differences between the calibration
results obtained by using our interpolation methods  and those
calibrated by assuming one particular cosmological model (the
$\Lambda$CDM model or the Riess cosmology ) with the same sample at
$z\le1.4$. This is not surprising, since the two  cosmological
models are consistent with the data of SNe Ia. Therefore, the GRB
luminosity relations calibrated from the cosmological models should
not be far from the true ones. However, we stress again that the
luminosity relations we obtained here are completely cosmological
model independent.

In order to constrain the cosmological parameters, we have applied
the calibrated relations to GRB data at high redshift. Since our
method does not depend on a particular cosmological model, when we
calibrate the parameters of GRB luminosity relations, the so-called
circularity problem can be completely avoided. We construct the GRB
Hubble diagram  and constrain cosmological parameters by the minimum
$\chi^2$ method as in SN Ia cosmology. From the 42 GRBs data
($z>1.4$) obtained by our interpolation method, we obtain
$\Omega_{\rm M}=0.25_{-0.05}^{+0.04}$ and
$\Omega_{\Lambda}=0.75_{-0.04}^{+0.05}$ for the flat $\Lambda$CDM
model, and for the dark energy model with a constant equation of
state $w_0=-1.05_{-0.40}^{+0.27}$ for a flat universe, which is
consistent with the concordance model within the statistical error.
Our result suggests the the concordance model ($w_0=-1$,
$\Omega_{\rm M}=0.27$, $\Omega_{\Lambda}=0.73$) which is mainly
derived from observations of SNe Ia at lower redshift is still
consistent with the GRB data at higher redshift up to $z=6.6$.

For the calibration of SNe Ia, the luminosity relations could evolve
with redshift, in such a way that local calibrations could introduce
biases (Astier et al., 2006; Schaefer, 2007). However, Riess et al.
(2007) failed to reveal the direct evidence for SN Ia evolution from
analysis of the $z>1$ sample-averaged spectrum. It is possible that
some unknown biases of SN Ia luminosity relations may propagate into
the interpolation results and thus the calibration results of GRB
relations. Nevertheless, the distance moduli of SN Ia are obtained
directly from observations, and therefore the observable distance
moduli of SNe Ia used in our interpolation method to calibrate the
relations of GRBs are completely independent of any given
cosmological model.

Recently, there have been discussions of possible evolution effects
and selection bias in GRB relations. Li (2007) used the Amati
relation as an example to test the possible cosmic evolution of GRBs
and found that the fitting parameters of the relation vary
systematically and significantly with the mean redshift of the GRBs.
However, Ghirlanda et al. (2008) found no sign of evolution with
redshift of the Amati relation and the instrumental selection
effects do not dominate for GRBs detected before the launch of the
\emph{Swift} satellite. Massaro et al. (2008) investigated the
cosmological relation between the GRB energy index and the redshift,
and presented a statistical analysis of the Amati relation searching
for possible functional biases. Tsutsui et al. (2008) reported a
redshift dependence of the lag-luminosity relation in 565 BASTE
GRBs. Oguri \& Takahashi
 (2006) and Schaefer (2007) discussed the gravitational lensing and
Malmquist biases of GRBs and found that the biases are small. Butler
et al. (2007) claimed that the best-fit of the Amati relation
presented in the \emph{Swift} sample is inconsistent with the
best-fit pre-\emph{Swift} relation at $>5 \sigma$ significance.
Butler et al. (2008) showed that the Ghirlanda relation is
effectively independent of the GRB redshifts. Nevertheless, further
examinations should be required for considering GRBs as standard
candles to cosmological use.

Because of the small GRB sample used to constrain the cosmology, the
present method provides no more accurate cosmological parameter
constraints than the previous works where the GRB and SN Ia samples
were used jointly. For example, Wang, Dai, \& Zhu (2007) presented
constraints on the cosmological parameters by combining the 69 GRB
data with the other cosmological probes to get $\Omega_{\rm
M}=0.27\pm0.02$. However, our main point is not on improvement of
the ``nominal statistical'' error; rather we emphasize that our
method avoids the circularity problem more clearly than previous
cosmology dependent calibration methods. Therefore, our results for
these GRB relations are less dependent on prior cosmological models.

\section*{Acknowledgements}
We thank Hao Wei, Pu-Xun~Wu, Rong-Jia~Yang, Zi-Gao~Dai, and En-Wei
Liang for kind help and discussions. We also thank the referee for
constructive suggestions and Yi-Zhong~Fan for valuable comments and
the mention of the reference (Takahashi K. et al. 2003). This
project was in part supported by the Ministry of Education of China,
Directional Research Project of the Chinese Academy of Sciences
under project KJCX2-YW-T03, and by the National Natural Science
Foundation of China under grants 10521001, 10733010, and 10725313.

\end{document}